\documentclass[10pt,conference]{IEEEtran}
\IEEEoverridecommandlockouts
\usepackage{cite}
\usepackage{stfloats} 
\usepackage{amsmath,amssymb,amsfonts}
\usepackage{algorithmic}
\usepackage[final]{graphicx} 
\usepackage{textcomp}
\usepackage{xcolor}
\usepackage{siunitx}
\usepackage{glossaries}
\usepackage{cleveref}
\usepackage{balance}
\Crefname{figure}{Fig.}{Figs.}
\usepackage{csquotes}
\usepackage{tikz}
\usetikzlibrary{arrows.meta,positioning,fit,calc,shapes.geometric}
\usepackage{pgfplots}
\usepackage{comment}
\newacronym{cccv}{CCCV}{constant-current constant-voltage}
\newacronym{ecm}{ECM}{equivalent-circuit model}
\newacronym{ev}{EV}{electric vehicle}
\newacronym{mpc}{MPC}{model predictive control}
\newacronym{ode}{ODE}{ordinary differential equation}
\newacronym{p2d}{P2D}{pseudo-two-dimensional}
\newacronym{pbm}{PBM}{physics-based model}
\newacronym{pde}{PDE}{partial differential equation}
\newacronym{pid}{PID}{proportional-integral-derivative}
\newacronym{sei}{SEI}{solid-electrolyte interphase}
\newacronym{soc}{SOC}{state-of-charge}
\newacronym{soh}{SOH}{state-of-health}
\newacronym{spm}{SPM}{single particle model}
\newacronym{spme}{SPMe}{single particle model with electrolyte dynamics}
\newacronym{ocv}{OCV}{open-circuit voltage}
\newacronym{ocp}{OCP}{open-circuit potential}
\newacronym{mimo}{MIMO}{multi-input multi-output}

\newcounter{todocounter}
\newcounter{ga}

\def\BibTeX{{\rm B\kern-.05em{\sc i\kern-.025em b}\kern-.08em
    T\kern-.1667em\lower.7ex\hbox{E}\kern-.125emX}}
\usepackage{pgfplots}
\pgfplotsset{compat=1.18}

\begin{document}

\title{Degradation-aware fast charging of Li-ion batteries using joint electrical and thermal model predictive control}

\author{Frederic Fabry\authorrefmark{1},
Alessio A. Lodge\authorrefmark{1},
Robinson Medina\authorrefmark{1},
Feye S.J. Hoekstra\authorrefmark{1},
Steven Wilkins\authorrefmark{1}\authorrefmark{2} and\\Madalin Frunzete\authorrefmark{3}
\thanks{\authorrefmark{1}Powertrains Dept., TNO Mobility and Built Environment, Helmond, The Netherlands. 
Emails: \{alessio.lodge, robinson.medina, feye.hoekstra, steven.wilkins\}@tno.nl (Note: frederic.fabry@gmx.de)}
\thanks{\authorrefmark{2}Eindhoven University of Technology, Department of Electrical Engineering, Eindhoven, The Netherlands. 
}
\thanks{\authorrefmark{3}National University of Science and Technology Politehnica Bucharest, Faculty of Electronics, Telecommunications and Information Technology, Bucharest, Romania. 
Email: madalin.frunzete@upb.ro}
\thanks{This research has received funding from the European Union through the Horizon 2020 research and innovation programme under grant agreement No 101103898 under title of NEXTBMS (https://nextbms.eu) and the Erasmus Mundus "Electric Vehicle Propulsion and Control" joint master program (https://master-epico.ec-nantes.fr/).}
}
\maketitle

\begin{abstract}
Fast charging of lithium-ion batteries is essential for electric vehicle adoption, but aggressive charging can accelerate its degradation and create safety risks. This study investigates a control framework that coordinates charging current with active thermal management to minimise charging time, while respecting constraints on electrochemical degradation and thermal safety. A single particle model with electrolyte dynamics (SPMe), extended with a two-node thermal model, represents the battery dynamics and enables the prediction of internal states - used in the control strategy - including anode potential, core temperature, and cell voltage.
Two multi-input multi-output control strategies are developed and compared: a classical approach using parallel proportional-integral-derivative (PID) controllers and an advanced model predictive control (MPC) with dual resolution prediction. Both controllers regulate the charging current and thermal resistance to minimise charging time while keeping within the limits of anode potential, core temperature, and cell voltage.
The results demonstrate that coordinated thermal-electrochemical optimal control outperforms conventional approaches, achieving a 42.2\% reduction in charging time compared to the manufacturer's charging recommendation, without increasing degradation. MPC, on average in the considered scenarios, reduces the charging time by 5.2\% compared to PID control, but at a significant computational cost.
\end{abstract}


\section{Introduction}
\Glspl{ev} are a key enabler of transport decarbonisation, but face significant barriers, with charging time remaining one of the primary challenges \cite{corradi_what_2023}. The most critical degradation mechanism in fast charging applications is lithium plating, which occurs when the anode potential drops below \SI{0}{\volt}. This causes the formation of lithium metal, which below 0 V is thermodynamically favourable over the intended intercalation reaction, which occurs during normal operation \cite{edge_lithium_2021}. Current fast charging approaches can be broadly categorised into heuristic and model-supported strategies \cite{wassiliadis_review_2021}. Heuristic methods such as \gls{cccv} protocols rely on predefined charging profiles with fixed current, voltage, or power constraints. Although simple to implement, these methods ignore real-time battery conditions and fail to optimise performance based on actual system dynamics. More sophisticated model-supported approaches have emerged using electrochemical models combined with control algorithms. These methods often utilise electrochemical models such as the \gls{p2d} model or reduced-order alternatives such as \gls{spme}, which provide detailed information on internal battery states while maintaining computational efficiency \cite{moura_battery_2017}.

Recent research has demonstrated the potential of electrochemical models for degradation-aware charging control. One of the few comparative studies between control approaches, comparing \gls{cccv}, \gls{pid}, and \gls{mpc} methods using a physics-enhanced equivalent circuit model was conducted in  \cite{li_electrochemical_2021}. Non-linear \gls{mpc} incorporating temperature optimisation and pulse discharging for lithium stripping achieved a 61\% reduction in charging time compared to 1C \gls{cccv} charging in \cite{yin_actively_2020}. Promising results using an \gls{spme} with \gls{pid} control of anode potential were shown in \cite{wassiliadis_model-based_2023}, achieving more than 1500 cycles with extensive validation studies that included detailed ageing analysis and post-mortem investigations.

Despite these advances, several critical gaps remain in the literature. First, the integration of active thermal management within health-aware model-based fast charging frameworks remains largely unexplored. Although temperature effects on battery degradation are well-established, most existing strategies treat thermal effects as constraints rather than actively leveraging thermal management as a control variable. Second, systematic comparison between classical and advanced control architectures using identical electrochemical models is limited. The computational requirements of advanced control methods like \gls{mpc} can be restrictive for real-time implementation, particularly when using high-fidelity electrochemical models, yet the performance benefits relative to simpler approaches remain unclear.

This work addresses these gaps by developing and fairly comparing, \textit{in silico}, degradation-aware model-based fast charging approaches that integrate electrochemical battery modelling with either \gls{pid} or \gls{mpc} control strategies, including direct regulation of battery temperature.

\section{Electrochemical battery model}
Accurate modelling of electrochemical and thermal dynamics is essential to design a fast and degradation-aware charging strategy. The \gls{spme} used in this work was originally derived in \cite{moura_battery_2017} and enhanced with a simple fast charging \gls{pid} controller in MATLAB in \cite{wassiliadis_model-based_2023}. A detailed validation of the model can be found in \cite{wassiliadis_systematic_2022}.

\subsection{Single particle model with electrolyte dynamics}
\label{subsec:spme}
In this paper, battery dynamics are represented using an \gls{spme}, which captures the essential electrochemical behaviour while being less computationally demanding compared to higher-fidelity models such as the \gls{p2d}. The \gls{spme} describes the transport of lithium-ions in both the solid electrode particles and the liquid electrolyte, allowing the prediction of internal states, which are critical for safe operation. For safe, degradation-aware control, three variables are of primary importance: the anode potential $\phi^-$, the cell core temperature $T_{c}$, and the cell terminal voltage $U_{c}$.

\subsubsection{Anode potential}
The anode potential represents the electrochemical potential at the negative electrode surface and serves as a key indicator for lithium plating prevention. It is calculated as
\begin{equation}
    \phi^-(t) = U_{OCP}^- \left( c_{ss}^-(t)\right) + \eta^{-}(t) + F R_{f}^- j_{n}^-(t),
    \label{eq:anode_potential}
\end{equation}
where $U_{OCP}^-$ is the open-circuit potential of the anode as a function of the surface concentration $c_{ss}^-$ (see eq. (9) in \cite{moura_battery_2017}), $\eta^-$ (see eq. (29) in \cite{moura_battery_2017}) is the activation overpotential, $F$ denotes the Faraday constant, $R_{f}^-$ is the film resistance of the \gls{sei} layer, and $j_{n}^-(t)$ describes the charge flux density (see eq. (22) in \cite{moura_battery_2017}). Lithium plating occurs when $\phi^-(t) < \SI{0}{\volt}$ vs. Li/Li$^+$, making this quantity critical for safe fast charging. Real-world applications typically use minimal margins above \SI{0}{\volt} to account for inhomogeneities and other localised phenomena.

\subsubsection{Cell voltage}
The cell voltage represents the potential difference between the cathode and anode
\begin{equation}
    U_{c}(t) = \phi^+(t) - \phi^-(t),
    \label{eq:cell_voltage}
\end{equation}
where $\phi^+(t)$ is the cathode potential (see eq. (28) in \cite{moura_battery_2017}).

\subsubsection{Thermal dynamics}
The \gls{spme} is augmented with a two-state lumped thermal model \cite{forgez_thermal_2010, perez_optimal_2017}:
\begin{subequations}\label{eq:thermal_model}
\begin{align}
    \frac{dT_{c}}{dt} &= \frac{T_{s}(t) - T_{c}(t)}{R_{in}C_{c}} + \frac{P_{gen}(t)}{C_{c}}, \label{eq:thermal_core}\\
    \frac{dT_{s}}{dt} &= \frac{T_{amb} - T_{s}(t) + P_{gen}(t) R_{in}}{C_{s}(R_{in} + R_{out}(t))}, \label{eq:thermal_surface}
\end{align}
\end{subequations}
where $T_{c}$ and $T_{s}$ represent core and surface temperatures, respectively, $R_{in}$ is the internal thermal resistance between the surface and the core, $C_{c}$ and $C_{s}$ are thermal capacitances in the core and surface, and $R_{out}(t)$ is the actuated external thermal resistance between the surface and the environment.
\begin{figure}
    \centering
    \includegraphics{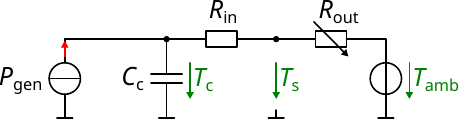}
    \caption{Lumped thermal model.}
    \label{fig:thermal_model}
\end{figure}
The heat generation rate is given by
\begin{equation}\label{eq:heat_generation}
    P_{gen}(t) = I(t)\left(U_{c}(t) - U_{{OCV}}(t)\right),
\end{equation}
which arises primarily from irreversible resistive effects during high-current operation \cite{oregan_thermal-electrochemical_2022}, where in this work $I(t)>0$~A denotes charging. Reversible entropic heating is assumed negligible in high current situations. The external thermal resistance $R_{out}$ effectively summarises the thermal resistance between the generic ambient conditions and the cell surface, thus mathematically representing different technological realisations of cooling systems with a constant sink temperature, such as natural or forced convection with a constant working fluid temperature.

For a comprehensive description of the underlying partial differential equations governing lithium concentration profiles and electrolyte dynamics, the reader is referred to \cite{moura_battery_2017, perez_optimal_2017, wassiliadis_systematic_2022}.

\subsection{Discrete-Time Representation for MPC}
\label{subsec:discretization}

The continuous-time \gls{spme} dynamics are discretised using a variable-order solver based on numerical differentiation formulas with sampling time $\Delta t = 1$ s, yielding the discrete-time state-space model
\begin{equation}\label{eq:discrete_dynamics}
    x_{k+1} = h(x_k, i_k, r_k),
\end{equation}
where $x_k \in \mathbb{R}^{n_x}$ contains the discretised lithium concentration, electrolyte and solid potential, and thermal states $[T_{c,k}, T_{s,k}]^\top$, with $T_{s,k} = T_{s} (k \Delta t)$, $T_{c,k} = T_{c} (k \Delta t)$ and $\Delta t$ and $k$ being the sampling period and discrete-time index, respectively. 
The control inputs are the applied current $i_k = i(k\Delta t)$ and external thermal resistance $r_k = R_{out}(k\Delta t)$. 

The states and outputs of interest for constraint enforcement are obtained through the mappings
\begin{equation}\label{eq:discrete_outputs}
    \begin{bmatrix}\phi^-_k\\ T_{{c},k}\end{bmatrix} = f(x_k, i_k, r_k), \qquad U_{{c},k} = g(x_k, i_k, r_k),
\end{equation}
where $f: \mathbb{R}^{n_x} \times \mathbb{R} \times \mathbb{R} \to \mathbb{R}^2$ evaluates the discretised version of \eqref{eq:anode_potential} and \eqref{eq:thermal_core}, and $g: \mathbb{R}^{n_x} \times \mathbb{R} \times \mathbb{R} \to \mathbb{R}$ evaluates \eqref{eq:cell_voltage}. This discrete-time representation forms the prediction model for the \gls{mpc} formulation presented in Section~\ref{sec:mpc}.

\subsection{System architectures}
The fast charging study by Wassiliadis et al. \cite{wassiliadis_model-based_2023} serves as a reference point. In the study, the charging current $i_k$ was controlled to track the reference value of the anode potential $\phi^-_k$, effectively creating a single-input single-output system. Since non-invasive measurement of anode potential is unfeasible, an open-loop model-based estimation (a virtual sensor) was implemented, with the plant model being the \gls{spme} mentioned above. The benefit of this type of control, based on anode potential actuation, is the combination of two objectives. First, reaching a low anode potential requires a large current, which automatically leads to minimising charging time. Secondly, utilising anode potential as reference limits degradation. As shown in \cite{wassiliadis_model-based_2023}, the anode potential limit does not need to be adjusted for different states of battery health.

Although this approach showcases the potential of using degradation-aware model-based fast charging strategies, it has shortcomings. Safe operation also requires limiting temperature and voltage in the cell, leading to an increased number of controlled outputs. Furthermore, additional system inputs may be considered. Thus, one aim of this work is to study the effects of thermal control on battery fast charging, limited to cooling with a constant ambient temperature. Inclusion of additional output constraints on temperature and voltage, together with the introduction of a second system input, results in a more complex \gls{mimo} system formulation which can be observed in \Cref{fig:PID_control_structure,fig:MPC_control_structure}. A detailed explanation of the respective controller designs is given in the following section. 

\section{Controller design}
Two control strategies are formulated to enforce safety constraints while minimising charging time. Classical \gls{pid} control is compared against a non-linear \gls{mpc} scheme, enabling direct comparison of their ability to coordinate charging current and thermal actuation under varying operating conditions.

\subsection{System characterisation}
\label{subsubs:System_charac}
\begin{figure}
    \centering
    \input{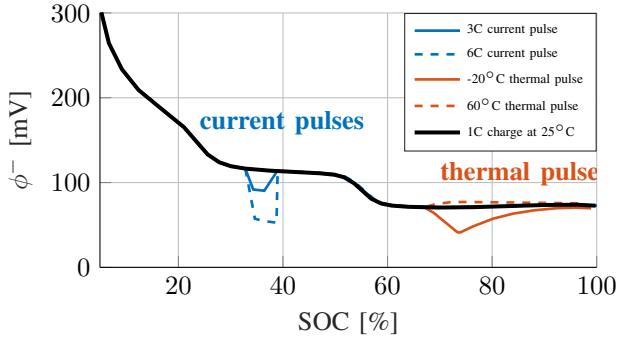}
    \caption{Anode potential evolution during a baseline charge cycle of 1C at 25 degrees. The `pulses' correspond to temporal deviations of the current or ambient temperature, on top of the baseline charge cycle.}
    \label{fig:sys_charac}
\end{figure}
Before a control strategy can be designed, the characteristics of the system must be derived. A common procedure is the measurement of the step function response. Since the final control strategies of \gls{pid} control and \gls{mpc} will actuate on two inputs, charging current $i_k$ and thermal resistance $r_k$, the influence of both inputs is investigated in \Cref{fig:sys_charac}.

The anode potential exhibits non-linear behaviour across different states of charge during the baseline 1C charge. Current step inputs at \SI{30}{\percent} \gls{soc} demonstrate rapid electrochemical response, with higher currents driving the anode potential closer to the critical \SI{0}{\volt} lithium plating boundary, highlighting degradation risks during fast charging. At \SI{70}{\percent} \gls{soc}, thermal step pulses are applied to the battery by abruptly changing the ambient temperature to \SI{-20}{\degreeCelsius} and \SI{60}{\degreeCelsius}, respectively, and adjust $r_k$ to forced convection. This reveals different system dynamics, where heating increases the anode potential (beneficial for fast charging applications), and cooling lowers the anode potential, leading to an increased risk of lithium plating. Thus, thermal management must balance the risks of low-temperature plating and the need to prevent excessive temperatures that compromise safety and accelerate thermal degradation mechanisms.
\subsection{PID design}
\label{subsubs:Modelling_Sys_architectures}
\begin{figure*}
    \centering
    \includegraphics{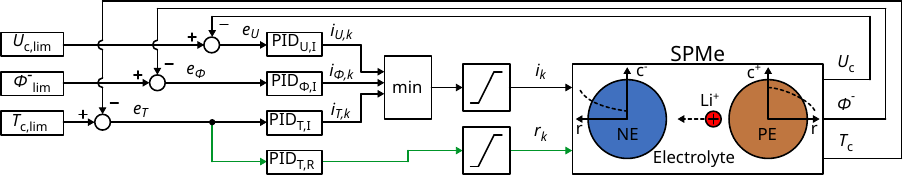}
    \caption{MIMO control architecture using PID control.}
    \label{fig:PID_control_structure}
\end{figure*}
\Cref{fig:PID_control_structure} shows how the classical control approach employs parallel \gls{pid} controllers for each constraint, with minimum-current logic for coordination. Each constraint generates a current input based on its individual \gls{pid} controller. The constraint errors are labelled $e_{U} = U_{c,lim} - U_{c}$, $e_{\phi} = \phi^- - \phi^-_{lim}$, and $e_{T} = T_{c,lim} - T_{c}$, with $U_{c,lim}$, $\phi^-_{lim}$ and $T_{c,lim}$ being the limits on cell voltage, anode potential and cell temperature, respectively. The final applied current is given by
\begin{equation}
i_k = \min( i_{U,k}, \ i_{\phi,k}, \ i_{T,k}).\label{eq:PIDCurrent}
\end{equation}
Thermal resistance is regulated using a separate \gls{pid} controller that targets optimal core temperature (green line in \Cref{fig:PID_control_structure}).

\subsection{MPC design}\label{sec:mpc}
\begin{figure*}
    \centering
    \includegraphics{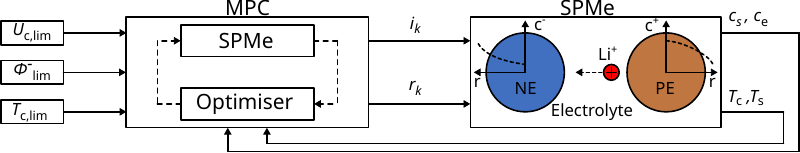}
    \caption{MIMO control architecture using MPC.}
    \label{fig:MPC_control_structure}
\end{figure*}
The \gls{mpc} structure uses a second \gls{spme} instance initialised with initial cell states $c_{s}, c_{e}$ (lithium concentrations) and $T_{c}, T_{s}$ to optimise control inputs over a prediction horizon $N$. The corresponding block diagram is shown in \Cref{fig:MPC_control_structure}. The associated optimisation problem is given by

\begin{subequations}\label{eq:optimization_problem}
\begin{align}
    \min_{i_{n|k}, r_{n|k}} \quad & \sum_{n=0}^{N-1} -\alpha \, i_{n|k}^2 + \beta \, (\Delta i_{n|k})^2 + \gamma \, (\Delta r_{n|k})^2 \label{eq:costFun} \\
    \text{s.t.} \quad 
    & x_{n+1|k} = h(x_{n|k}, i_{n|k}, r_{n|k}), \quad x_{0|k} = x_k \label{eq:state_dynamics}\\
    & \begin{bmatrix}\phi^-_{n|k}\\ T_{{c},n|k}\end{bmatrix} = f(x_{n|k}, i_{n|k}, r_{n|k}), \label{eq:outputs_f}\\
    & U_{{c},n|k} = g(x_{n|k}, i_{n|k}, r_{n|k}), \label{eq:outputs_g}\\ 
    & I_{min} \leq i_{n|k} \leq I_{max},\label{eq:input_bounds_current}\\
    & R_{out,min} \leq r_{n|k} \leq R_{out,max},\label{eq:input_bounds_therm_res}\\
    & U_{{c},n|k} \leq U_{c,lim},\label{eq:output_constraint_volt}\\
    & \phi_{lim}^- \leq \phi^-_{n|k}, \label{eq:output_constraint_potn}\\
    & T_{{c},n|k} \leq T_{c,lim}. \label{eq:output_constraint_temp}
\end{align}
\end{subequations}
where $n|k$ denotes the prediction at time $k + n$,  $n \in \{0, \ldots, N-1\}$ denotes the prediction step with $N$ the prediction horizon , $k\in\{1,\dots,K\}$, $x_{0|k}$ are the initial conditions with $K$ being the final simulation time, and subscripts $_{min}$, $_{max}$ denote minimum and maximum values of the various controlled inputs, states and outputs. The cost function promotes fast charging through the first term, where $i_{n|k}>0$ denotes charging, while the terms $\Delta i_{n|k} = i_{n-1|k} - i_{n|k}$ and $\Delta r_{n|k} = r_{n-1|k} - r_{n|k}$ penalise rapid changes in control inputs, ensuring smooth actuation. The weighting factors $\alpha, \beta, \gamma > 0$ balance between the objectives of current delivery (first term) and smooth control trajectories (second and third terms). The constraints \eqref{eq:input_bounds_current}-\eqref{eq:input_bounds_therm_res} reflect the limitations of the charger and thermal system. Safety constraints \eqref{eq:output_constraint_volt}-\eqref{eq:output_constraint_temp} prevent overcharging, lithium plating, and thermal degradation, respectively. At each time step $k$, the optimiser solves for the optimal inputs $i_{n|k}, r_{n|k}$ and applies the first control action. The optimiser uses a warm-start initialisation of the previous solution to reduce the convergence time.

The prediction horizon employs a dual resolution strategy to efficiently handle both rapid and gradual system dynamics. From the simulations run, electrochemical responses, particularly anode potential and cell voltage variations, are fast and require fine short-term temporal resolution, whereas thermal dynamics are slower and can be adequately represented with coarser long-term time steps. 
The total prediction horizon spans $N = N_{fine} + N_{coarse}$ steps with variable time step size
\begin{equation}
    \Delta t_n =
    \begin{cases}
        \Delta t_{fine}, & n \in \{ 0,\ldots,N_{fine}-1\}, \\
        \Delta t_{coarse}, & n \in \{N_{fine},\ldots,N_{fine}+N_{coarse}-1\},
    \end{cases}
    \label{eq:pred_horizon_implementation}
\end{equation}
where $\Delta t_{fine} < \Delta t_{coarse}$. This segmented approach reduces the dimensionality of the optimisation problem, thereby decreasing computational requirements.

\section{Methodology}
Having established the theoretical foundation for both \gls{pid} and \gls{mpc} control strategies, the practical implementation of these approaches requires careful consideration of computational aspects, software tools, and simulation parameters. The following section details the technical implementation decisions that enable effective comparison between the two control architectures under realistic operating conditions. Both control approaches were developed using MATLAB R2024b, leveraging constrained optimisation with \texttt{fmincon} function. The \gls{pde} are solved using MATLAB's \texttt{ode15s} function. The control system builds upon the foundational \gls{spme} model developed by Wassiliadis et al. \cite{wassiliadis_systematic_2022}, which was enhanced to incorporate thermal management and \gls{mimo} system capabilities. The simulations were performed on a laptop equipped with an Intel Core 1.3 GHz i5-1335U processor with 16 GB of RAM memory. The simulated battery is a graphite|NCA Sony/Murata 18650 (US18650VTC5A) cell, with a nominal capacity of \SI{2.5}{Ah} and a maximum charging current of \SI{6}{\ampere} at \SI{20}{\degreeCelsius}, specified by the manufacturer.

The \gls{spme} is spatially discretised with 20 nodes along the radial direction of the particles, and 20 nodes distributed along the x-direction, which cover the length of the cell including both electrodes and the separator. For the \gls{mpc} approach, a Sequential Quadratic Programming (SQP) algorithm serves as the optimisation method within \texttt{fmincon}, adhering to MATLAB's recommended practices for non-linear constrained optimisation. All optimisation settings maintain the default values except for the step tolerance, which is configured to $10^{-5}$. 

Both controllers operate at a base sampling interval of \SI{1}{\second}. The \gls{mpc} employs a dual-resolution prediction horizon, discretised at \SI{1}{\second} for the fast electrochemical dynamics and \SI{15}{\second} for the slower thermal dynamics, whereas the \gls{pid} controller applies a uniform \SI{1}{\second} interval throughout. Control input limitations are enforced, with the charging current capped at \SI{15}{\ampere} (6C) to maintain consistency with the model validation boundaries established in \cite{wassiliadis_systematic_2022}. Research by Steinhardt et al. \cite{steinhardt_low-effort_2021} determined the thermal resistance $R_{out}$ to be \SI{14}{\kelvin\per\watt} under natural convection conditions with \SI{0.6}{\meter\per\second} airflow around an 18650 cell. Under forced convection with \SI{3.1}{\meter\per\second} airflow, the thermal resistance reduces to \SI{4.2}{\kelvin\per\watt}.The \gls{pid} control implementation requires distinct parameter sets for proportional ($K_{P}$), integral ($K_{I}$), and derivative ($K_{D}$) gains across multiple controllers. For the \gls{mimo} configuration, the parameters are given in \Cref{tab:PID_gains}. Given the varying dynamics of non-linear systems across different operating regions, conventional empirical tuning approaches like Ziegler-Nichols have limited applicability, requiring a trial-and-error approach. All \gls{pid} controllers incorporate anti-windup mechanisms to address the current limitation enforced by the parallel \gls{pid} controllers.
\begin{table}
    \centering
    \begin{tabular}{c|c|c|c|c}
         & ${PID}_{U,I}$ & ${PID}_{{\phi,I}}$ & ${PID}_{T,I}$ & ${PID}_{T,R}$ \\
         \hline
        $K_{P}$ & 0.05 & 0.05 & 30 & 10\\
        $K_{I}$ & 0 & 0.05 & 0 & 0\\
        $K_{D}$ & 3 & 0.001 & 0 & 1\\
    \end{tabular}
    \caption{PID controller gains.}
    \label{tab:PID_gains}
\end{table}

For the \gls{mpc} implementation, the current maximisation term dominates the cost function and is normalised to unity by setting $\alpha=\frac{1}{I_{max}^2N}=0.55 \times 10^{-3}$. Control smoothness penalties are weighted with $\beta=10$ and $\gamma=1\times 10^{-3}$. Input constraints match those specified for the \gls{pid} implementation. The horizon is evenly split with $N_{fine}=N_{coarse}=4$. Time discretisation is selected as the simulation sampling interval of $\Delta t_{fine}=\SI{1}{\second}$ and $\Delta t_{coarse}=\SI{15}{\second}$. Safety and reliability during fast charging are maintained through electrochemical and thermal constraint enforcement. The lower bound of the anode potential is established at $\phi_{lim}^- = \SI{45}{\milli\volt}$ (derivation available in \cite{wassiliadis_model-based_2023}). The upper voltage limit $U_{c,lim}$ is cell dependent and equals \SI{4.2}{\volt} for the cell studied. These constraints apply only to the fine dynamics segment $n \in \{0,\dots, N_{fine}-1\}$ to reduce computational complexity. The upper temperature bound is set at $T_{c,lim} = \SI{45}{\degreeCelsius}$, based on observations by Wassiliadis et al. regarding the onset of non-linear ageing above this threshold \cite{wassiliadis_model-based_2023}. This constraint covers the entire prediction horizon $n \in \{0,\dots, N_{fine}+N_{coarse}-1\}$.

\section{Results}
\begin{figure}
    \centering
    \input{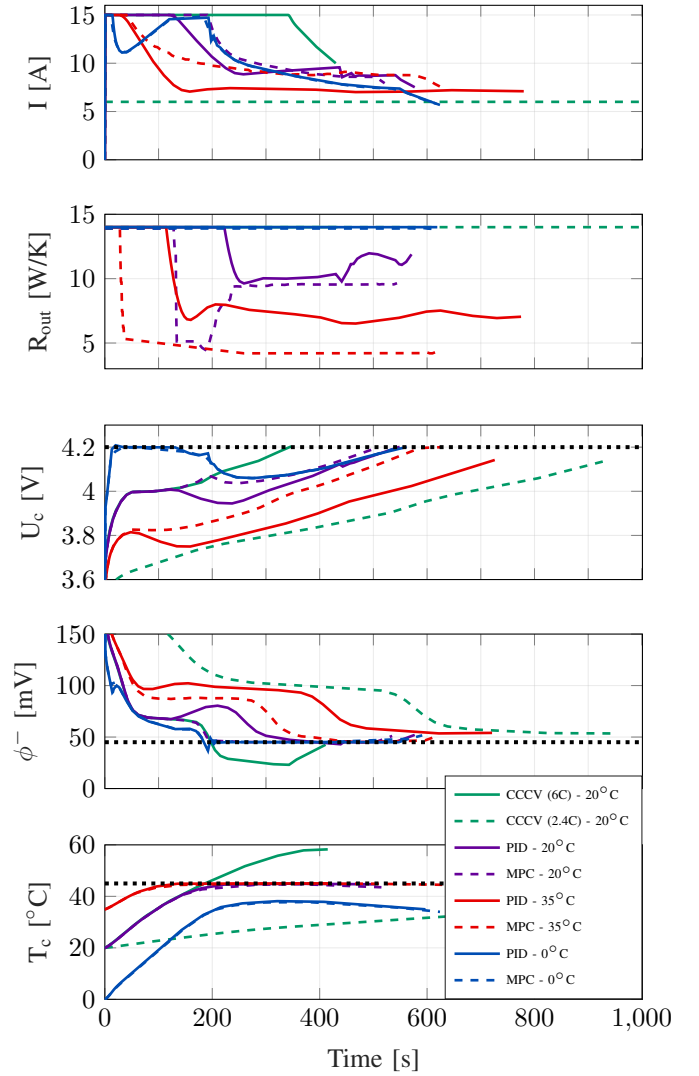}
    \caption{Control method comparison of CCCV, PID and MPC.}
    \label{fig:ChargComp}
\end{figure}
In this section, the simulation results demonstrate the performance characteristics of both control strategies across three thermal scenarios and one degradation case. For comparison, a standard \gls{cccv} charging strategy is given, with a charging current of \SI{6}{\ampere} (2.4C, manufacturer specification) and \SI{15}{\ampere} (6C, model validation boundary). The green charging profiles in \Cref{fig:ChargComp} showcase the \gls{cccv} approach, starting with a constant current phase until the specified voltage is reached and subsequently kept by reducing the current. When applying the charging profile recommended by the battery manufacturer (green dashed line), none of the degradation and safety limits are violated. However, this comes at the cost of a longer charging time (16.6 min). Using \gls{cccv} with a high current limit results in a violation of the anode potential and temperature limit (see solid green line). 

In \Cref{fig:ChargComp}, the performance of the \gls{pid} and \gls{mpc} is shown, for ambient (\SI{20}{\degreeCelsius}), cold (\SI{0}{\degreeCelsius}), and hot (\SI{35}{\degreeCelsius}) conditions, with all charging sessions spanning a \SI{10}{\percent} to \SI{80}{\percent} \gls{soc} window, representing a deep charging cycle for \glspl{ev}. At \SI{20}{\degreeCelsius}, both controllers maintain constraints on $U_{c}$, $\phi^-$ and $T_{c}$. \Gls{mpc} activates thermal management \SI{100}{\second} earlier than \gls{pid}, achieving \SI{30}{\second} faster charging (\SI{5.2}{\percent}) by maintaining a higher average current near the temperature limits. 

Cold conditions (\SI{0}{\degreeCelsius}) make anode potential and voltage the dominant constraints. Both controllers reduce the current to prevent $U_{c}$ exceeding \SI{4.2}{\volt}, with $\phi^-$ becoming limiting after \SI{190}{\second}. The thermal management maintains a higher $R_{out}$ to retain heat. \Gls{mpc} avoids the anode potential limit violation of the \gls{pid} controller observed at \SI{190}{\second}. Both controllers maintain voltage compliance through current reduction beyond \SI{550}{\second}.

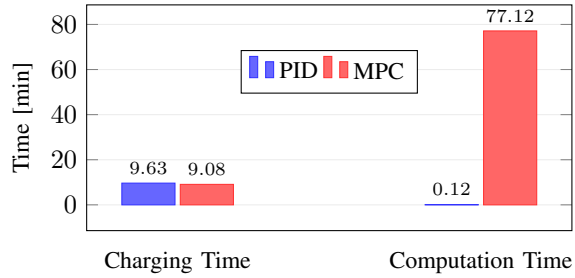
\begin{figure}
\centering
    \begin{tikzpicture}
        \begin{axis}[
            width=0.45\textwidth,
            height=0.19\textheight,
            ylabel={Time [min]},
            symbolic x coords={Charging Time, Computation Time},
            xtick=data,
            xtick style={draw=none},
            x tick label style={font=\small},
            ylabel style={font=\small},
            enlarge x limits=0.3,
            legend style={
                at={(0.5,0.8)},
                anchor=north,
                legend columns=-1,
                font=\small
            },
            ybar=2pt,
            bar width=20pt,
            ymajorgrids=true,
            grid style={opacity=0.3},
            enlarge y limits={upper,0.15},
        ]
        
        \addplot[
            fill=blue!60,
            draw=blue!80,
            nodes near coords,
            nodes near coords style={font=\scriptsize, above},
        ] coordinates {
            (Charging Time,9.63)
            (Computation Time,0.12)
        };
        
        \addplot[
            fill=red!60,
            draw=red!80,
            nodes near coords,
            nodes near coords style={font=\scriptsize, above},
        ] coordinates {
            (Charging Time,9.08) 
            (Computation Time,77.12) 
        };
        
        \legend{PID, MPC}
        \end{axis}
    \end{tikzpicture}
\caption{Comparison of charging time and computation time for PID and MPC control strategies for a \SI{20}{\degreeCelsius} charging session.}
\label{fig:BarGraph_charging_comparison}
\end{figure}

At elevated temperature (\SI{35}{\degreeCelsius}), the thermal constraints are activated early. \Gls{mpc} proactively selects low $R_{out}$ (forced convection) to sustain higher current without exceeding $T_{c}$ limits throughout the cycle. \Gls{pid} responds reactively, cooling only near temperature limits without full utilisation of thermal actuation. This results in a \SI{19.5}{\percent} longer charging time for \gls{pid} (\SI{780}{\second} versus \SI{628}{\second}).
The \gls{mpc} requires 77.12 minutes computation time versus 0.12 minutes for \gls{pid} (see \Cref{fig:BarGraph_charging_comparison}). The mean computation time per instance is \SI{0.012}{\second} for the \Gls{pid} controller and \SI{8.49}{\second} for the \gls{mpc}, corresponding to a factor of approximately 700 in computational overhead.

\begin{figure}
    \centering
    \input{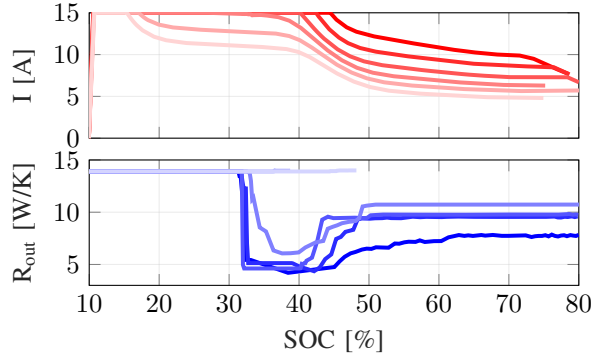}
    \caption{Evolution of the fast charging strategy with current and thermal management under MPC control. Ageing effects are represented by a reduction in capacity from \SI{100}{\percent} (solid line) to \SI{80}{\percent} (faded line), accompanied by a progressive increase in \gls{sei} film resistance $R_{f}^-$ from its initial value to twice that value. All charging cycles were simulated at \SI{20}{\degreeCelsius}.}
    \label{fig:ChargComp_agedCell}
\end{figure}

\Cref{fig:ChargComp_agedCell} demonstrates \gls{mpc}'s adaptation to varying \gls{soh} through increased \gls{sei} resistance and capacity reduction. Fresh cells permit higher currents with larger anode potential margins, generating more heat which requires active cooling. Aged cells reach potential limits earlier given the higher SEI film resistance, requiring smaller currents, thus producing less heat and eliminating cooling needs. The reduction in current occurs progressively earlier with degradation, with the most aged cell de-rating at \SI{15}{\percent} \gls{soc}, resulting in increased charging times. Although the results presented in the section are promising, real-world hardware in the loop testing should be developed, as the current study assumes perfect model match and parameterisation.

\section{Conclusion}
This work presents an electrochemically informed control framework for degradation-aware fast charging, combining current and thermal management through coordinated control inputs. The approach employs an extended \gls{spme} with thermal dynamics and compares classical \gls{pid} control against advanced \gls{mpc} with dual resolution prediction to capture both fast electrochemical and slow thermal dynamics.

The results demonstrate that thermal actuation effectively unlocks current headroom by preventing temperature violations, enabling \gls{mpc} to achieve \SI{5.2}{\percent} charging time reduction through predictive coordination of control inputs. Although \gls{mpc} provides superior performance through earlier and smoother control actuation, this comes at significant computational cost, with charge simulation times of 77.12 minutes compared to 0.12 minutes for \gls{pid}. The controller successfully adapts to varying cell health states, with aged cells requiring progressively earlier current reduction to maintain safety constraints.

Future work should focus on computational efficiency through reduced-order models and tailored optimisation routines to increase practical feasibility. Cycle life improvements should be experimentally verified with ageing studies. Real-life applications would require embedded deployment and an observer-in-the-loop which would increase computational demand for both \gls{pid} control and \gls{mpc}. Enhanced thermal management systems that enable both heating and cooling capabilities across diverse operating conditions could unlock the potential of predictive control further.
\balance
\bibliographystyle{IEEEtran}
\bibliography{references}

\end{document}